\documentclass{aastex}			

\usepackage{emulateapj5}	
\slugcomment{Published in the October 10 issue of ApJL.}

\shorttitle{MIR identification of faint SMM sources}
\shortauthors{Y.~Sato et al.}

\received{2001 July 31}
\begin{document}

\title{Mid-Infrared Identification
of Faint Submillimeter Sources\altaffilmark{1}}

\author{Y.~Sato\altaffilmark{2},
        L.L.~Cowie\altaffilmark{3},
        K.~Kawara\altaffilmark{2},
        Yoshiaki~Taniguchi\altaffilmark{3,4},
        Y.~Sofue\altaffilmark{2},
        H.~Matsuhara\altaffilmark{5},
        and
        H.~Okuda\altaffilmark{6}
        }

\altaffiltext{1}{
Based on observations with ISO,
an ESA project with instruments funded by ESA Member States
(especially the PI countries: France, Germany, the Netherlands
and the United Kingdom) and with the participation of ISAS and NASA.
}
\altaffiltext{2}{
Institute of Astronomy, University of Tokyo,
2--21--1 Osawa, Mitaka, Tokyo, 181--0015 Japan;
\tt{ysato@ioa.s.u-tokyo.ac.jp}
}
\altaffiltext{3}{
Institute for Astronomy, University of Hawaii,
2680 Woodlawn Drive, Honolulu, HI 96822
}
\altaffiltext{4}{
Astronomical Institute, Graduate School of Science, Tohoku University,
Aramaki, Aoba, Sendai, 980--8578 Japan
}
\altaffiltext{5}{
Institute of Space and Astronautical Science (ISAS),
3--1--1 Yoshinodai, Sagamihara, Kanagawa, 229--8510 Japan
}
\altaffiltext{6}{
Gunma Astronomical Observatory,
6860--86 Nakayama, Takayama, Agatsuma, Gunma, 377--0702 Japan
}

\begin{abstract}

Faint submillimeter sources
detected with the Submillimeter Common-User Bolometer Array
on the James Clerk Maxwell Telescope
have faced an identification problem
due to the telescope's broad beam profile.
Here we propose a new method to identify such submillimeter sources
with a mid-infrared image having a finer point spread function.
The Infrared Space Observatory has provided
a very deep 6.7\,$\mu$m image of the Hawaii Deep Field SSA13.
All three faint 850\,$\mu$m sources in this field
have their 6.7\,$\mu$m counterparts.
They have been identified with interacting galaxy pairs in optical images.
These pairs are also detected in the radio.
Two of them are optically faint and very red ($I>24$, $I-K>4$),
one of which has a hard X-ray detection with the $Chandra$ satellite.
As these observing properties are similar to
those of local ultraluminous infrared galaxies,
their photometric redshifts are derived
based on submillimeter to mid-infrared flux ratios
assuming a spectral energy distribution (SED) of Arp220.
Other photometric redshifts are obtained
via $\chi^2$ minimization between the available photometry data
and template SEDs.
Both estimates are in the range $z=1$--2,
in good agreement with a spectroscopic redshift and a millimetric one.
The reconstructed Arp220 SEDs with these redshift estimates 
are consistent with all the photometry data
except $Chandra$'s hard X-ray detection.
The sources would be a few times more luminous than Arp220.
With an assumption that AGN contributions are negligible,
it appears that extremely high star formation rates
are occurring in galaxies at high redshifts
with massive stellar contents already in place.

\end{abstract}

\keywords{
cosmology: observations ---
galaxies: evolution ---
galaxies: formation ---
infrared: galaxies ---
submillimeter
}

\objectname[\[BCS99\] J131232.1+424430]{}
\objectname[\[BCS99\] J131228.0+424458]{}
\objectname[\[BCS99\] J131225.7+424350]{}

\section{Introduction}

The installation of the Submillimeter Common-User Bolometer Array (SCUBA)
on the 15\,m James Clerk Maxwell Telescope
has brought us the discovery of a new population of faint submillimeter sources
\citep{SIB97,BCS+98,HSD+98,ELG+99}.
Meanwhile, the far-infrared background (FIRB) was detected
with the Far Infrared Absolute Spectrometer (FIRAS)
and the Diffuse Infrared Background Experiment (DIRBE)
on the COBE satellite
(\citealp{PAB+96}; \citealp*{SFD98}; \citealp{HAK+98,FDM+98}).
The FIRB is comparable to
or larger than that in the optical wavelengths,
and much of it could be explained
by integrating individual submillimeter sources down to the faintest limit
\citep{BKI+99}.
Thus, it is indispensable to investigate the faint submillimeter population
in order to understand overall history of energy production in the Universe.

Although SCUBA has unveiled the faint submillimeter population,
its beam is too broad (15{\arcsec} FWHM at 850\,$\mu$m)
to pinpoint their optical counterparts
\citep*[e.g.][]{BCR00}.
The high density of the submillimeter sources
also raises a problem of source confusion
\citep{ELW+00,H01}.
Up until now, methods to bypass the broad submillimeter beam
have included
centimetric radio interferometry with VLA
\citep{ISB+98,R99,SIK+99,BCR00}
and millimetric interferometry with IRAM or OVRO
\citep{DNG+99,GLS+00,BCM+00,FSIS00}.
Most recently,
an attempt to observe at a longer wavelength
has started at the IRAM 30\,m telescope
\citep{BCM+00}.
The larger dish provides
a slightly smaller beam (10{\arcsec} at 1.3\,mm).

Here we investigate the identification of SCUBA sources
with a mid-infrared deep image having a relatively small beam
(7\farcs2 at 6.7\,$\mu$m).
Faint submillimeter sources, presumably dusty systems at high redshifts,
could be bright in the mid-infrared regardless of their energy sources.
If the submillimeter emission was produced
by dust heated by star forming activity,
there should be stellar systems
following the stellar initial mass function.
The stellar systems could be luminous
at the rest-frame near-infrared ($>1\,\mu$m)
even in the presence of dust.
In the case that the source was powered by an active galactic nucleus (AGN),
very hot dust surrounding the AGN could emit its reprocessed energy
at $>2\,\mu$m.
For both cases,
the emitted light could be received at the mid-infrared,
even though the sources were at high redshifts.

\section{Source Identification}

\begin{figure*} 
  \plotone{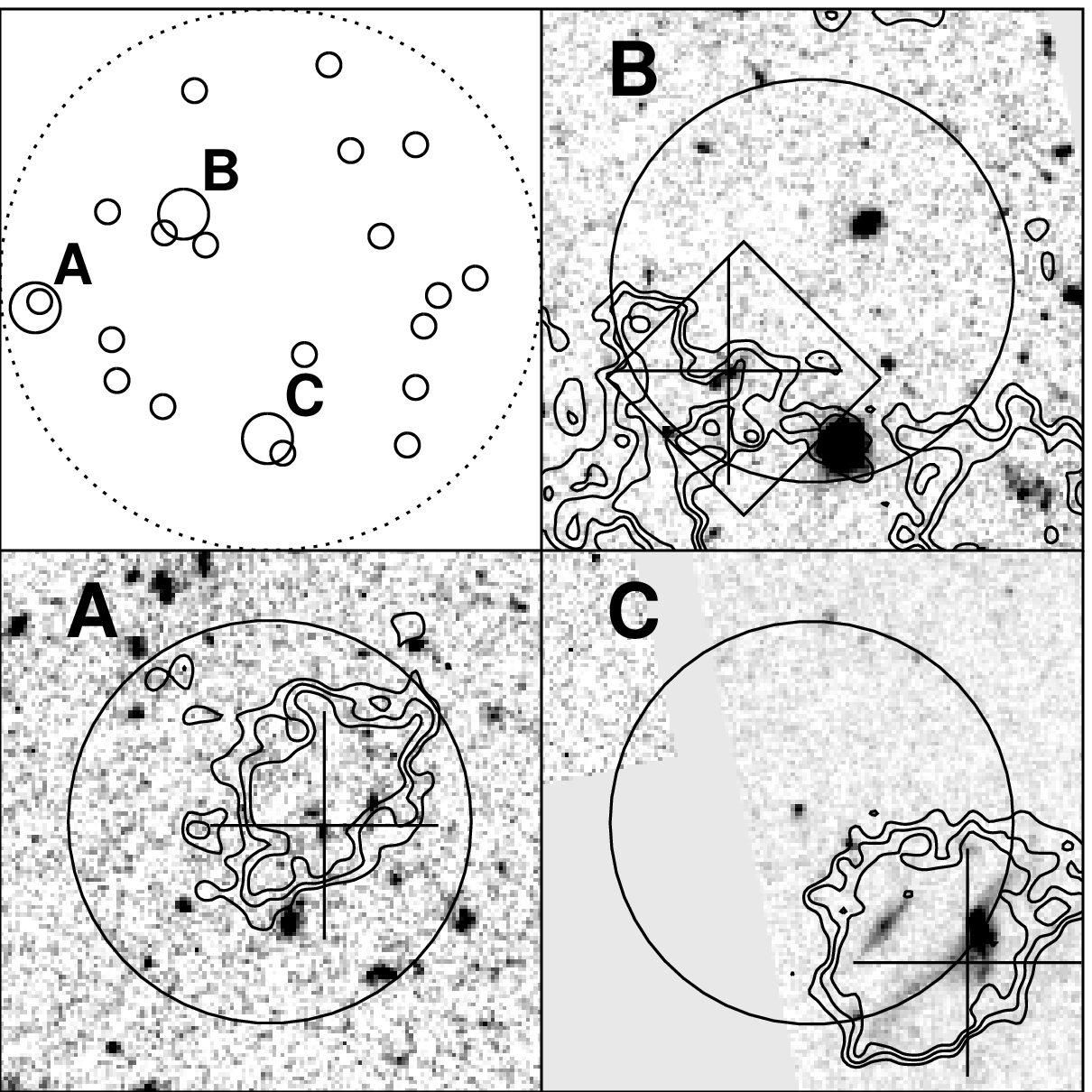}
  \caption{
Identification of submillimeter sources
A, B, and C (IDs in Table~\ref{tab:id}).
In the upper left panel,
the spatial distribution of the submillimeter and mid-infrared sources
is shown
in the SCUBA SSA13 deep field (2\farcm7 in diameter; dotted circle).
North is up and east to the left.
SCUBA 850\,$\mu$m and ISOCAM 6.7\,$\mu$m sources
\citep{BCS99,SKC+01}
are marked with large and small circles
with their beam sizes (15{\arcsec} and $7\farcs2$ FWHM, respectively).
Magnified views of the submillimeter positions are displayed on
the $HST$/WFPC2 $I$ band images (20{\arcsec} width in gray scale).
The 6.7\,$\mu$m circles are substituted with contours
to make full use of the smaller beam size and higher significance
of the 6.7\,$\mu$m sources
than those of the 850\,$\mu$m detections.
Contour levels are (2.5, 3.0, and 3.5) $\sigma$,
while (2.0, 2.5, and 3.0) $\sigma$ for source B.
Three VLA 1.4\,GHz sources and one $Chandra$ 2--10\,keV source
are shown with crosses and a diamond, respectively
(E.\ A.\ Richards 2001, private communication; \citealp{MCB+00}).
  }
\label{fig:id}
\end{figure*}

A sensitive mid-infrared camera ISOCAM
\citep{CAA+96}
on board the Infrared Space Observatory
\citep[ISO,][]{KSA+96}
provided us with
the first opportunity to identify faint submillimeter sources
at the mid-infrared.
In the Hawaii Deep Field SSA13,
a very deep mid-infrared map was obtained
with the LW2 (6.7\,$\mu$m) filter
(\citealt{SKC+01}).
This field was also surveyed extensively with SCUBA
(\citealp{BCS+98}; \citealp*{BCS99}).
In the upper left panel of Fig.~\ref{fig:id},
we plot both 6.7\,$\mu$m and 850\,$\mu$m sources
detected within the SCUBA field-of-view.
All three 850\,$\mu$m sources; A, B, and C
(these identifiers are used throughout this paper; Table~\ref{tab:id})
are found to be located very close to certain 6.7\,$\mu$m sources.
Monte Carlo simulations of source detection
suggest that displacement of the detected position
as large as the beam size
could happen in low signal-to-noise ratio (S/N) ranges
\citep{ELW+00,H01}.
We estimate the probability of random alignments
between the 6.7\,$\mu$m and 850\,$\mu$m sources.
As the surface density of 6.7\,$\mu$m sources at this depth
is larger than that of 850\,$\mu$m sources,
we assume a Poisson distribution for 6.7\,$\mu$m sources.
Then, the chance alignment probability can be defined as
$P = \exp ( - \pi r^2 N(6.7\,\mu\mathrm{m}))$,
where $r$ is the distance between 6.7\,$\mu$m and 850\,$\mu$m sources
and $N(6.7\,\mu\mathrm{m})$ is a 6.7\,$\mu$m integral number count
at the flux of the corresponding 6.7\,$\mu$m source.
The chance probability became only $P=0.9$, 16, and 4.1\,\%
for sources A, B, and C, respectively.
Then, we associate
the closest 6.7\,$\mu$m sources
with the corresponding submillimeter sources.

The optical counterparts of the sources were identified
in deep $HST$ images
taken with the WFPC2 F814W filter
\citep{CHS95}.
Interacting galaxy pairs were found
at the peaks of the 6.7\,$\mu$m contours
for all the 850\,$\mu$m sources
(Fig.~\ref{fig:id}).
The occurrence of an irregular morphology is consistent
with the results of ultraluminous infrared galaxies (ULIGs)
discovered with IRAS
\citep{SM96}.
Within 1{\arcsec} from all the optical counterparts,
faint 1.4\,GHz sources are detected
(E.\ A.\ Richards 2001, private communication).
Source B
also has a hard X-ray (2--10\,keV) detection with $Chandra$
\citep{MCB+00}.
The small rate of AGNs in submillimeter sources (1/3)
is similar to the previous results
\citep{ALB99,FSI+00,HBG+00,SMS+00,BCMR01}.
Magnitudes at the $K$, $I$, and $B$ bands were measured
with additional data taken from the ground
\citep{CSH+96}.
Two of the three sources are optically faint and very red
(Table~\ref{tab:id}).
Here are notes for the individual sources.
{\bf A)}
This is the most significant SCUBA detection in this sample
($4.7\,\sigma$).
The ISOCAM contour reduces
the uncertainty of the SCUBA position.
In the middle of the 6.7\,$\mu$m contour, 
there is a faint ($I \sim 25$) interacting galaxy pair
having a very red color of $I-K>4$.
{\bf B)}
One of the 6.7\,$\mu$m peaks
matches a colliding galaxy pair with $I \sim 24$ and $I-K>4$.
The ISOCAM source appears somewhat extended
probably due to confusion of faint sources.
{\bf C)}
A bright merger ($I \sim 22$) with tidal tails
is located at the edge of the beam centered at the nominal SCUBA position.
The original SCUBA image shows a sign of elongation,
suggesting a lower position accuracy
\citep[Fig.~1 in][]{BCS+98}.

\section{Redshift Estimates}

\begin{figure*}
  \plotone{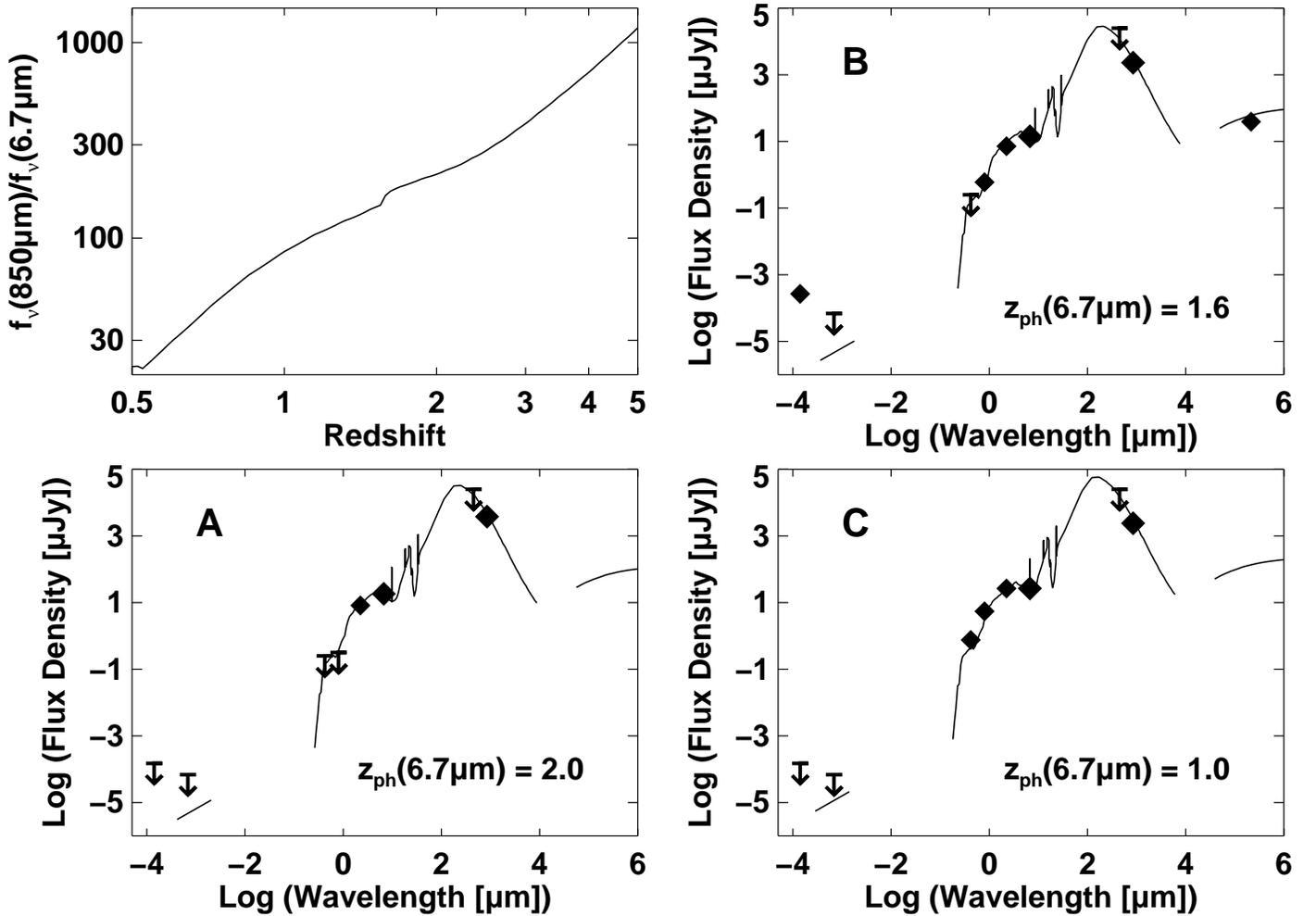}
  \caption{
Submillimeter to mid-infrared flux ratios
$f_\nu$(850\,$\mu$m)/$f_\nu$(6.7\,$\mu$m)
for an Arp220 SED,
showing a monotonic increase with redshift
(upper left panel).
Based on this relation,
photometric redshifts $z_{\rm ph}$(6.7\,$\mu$m)
are derived for the three submillimeter sources
A, B, and C.
With these redshift estimates,
Arp220 SEDs are reconstructed to fit
the 850\,$\mu$m and 6.7\,$\mu$m measurements.
We adopt the GRASIL SED
\citep{SGB+98},
complemented with radio and X-ray data
\citep{CY00,I99}.
Reconstructed SEDs are consistent with other photometry data.
The hard X-ray detection of source B is beyond the wavelength span of the SED.
The fluxes and upper limits (3\,$\sigma$) are measurements at
2--10\,keV, 0.5--2\,keV, $B$, $I$, $K$, 6.7\,$\mu$m, 
450\,$\mu$m, 850\,$\mu$m, and 1.4\,GHz
(Table~\ref{tab:id}; \citealp{MCB+00}; \citealp{BCS+98}).
  }
\label{fig:redshift}
\end{figure*}

The three SCUBA sources detected at 6.7\,$\mu$m
share the same observational properties
as those of local ULIGs;
i.e. an irregular morphology,
a red color, and optical and X-ray faintness.
Therefore it seems possible to estimate their redshifts
with an assumption
that their spectral energy distributions (SEDs)
are similar to that of Arp220,
an archetypal ULIG in the local universe.
We adopt a UV to submillimeter SED
in the GRASIL library to represent Arp220
\citep{SGB+98}.
Applying a convolution with two bandpass filters
at 850\,$\mu$m and 6.7\,$\mu$m,
we derived flux ratios
$f_\nu$(850\,$\mu$m)/$f_\nu$(6.7\,$\mu$m)
as a function of redshift
(upper left panel of Fig.~\ref{fig:redshift}).
Because of a monotonic increase with redshift,
this relation can be used to estimate redshifts,
analogous to the redshift estimator
based on a submillimeter to radio flux ratio
\citep{CY99,CY00,BCR00}.
The derived estimates were in the range
$z_{\mathrm{ph}}(6.7\,\mu\mathrm{m})=1$--2
(Table~\ref{tab:z}).
The reconstructed SEDs are consistent with other photometry data,
suggesting the usefulness of this redshift estimator
(Fig.~\ref{fig:redshift}).

With the identification results above,
we could omit the assumption of an Arp220 SED to obtain photometric redshifts.
Other estimates $z_{\rm ph}$(SED)
were derived via $\chi^2$ minimization
between all the photometry data and SED templates.
We here adopt the GRASIL SED library
\citep{SGB+98};
UV to submillimeter SEDs of
local templates (starbursts: M82, NGC6090, and Arp220,
normal galaxies: M51, M100, and NGC6946, and a giant elliptical)
and UV to radio SEDs of evolving E, Sa, Sb, and Sc galaxies.
We add an X-ray to radio SED of a heavily obscured AGN NGC6240 compiled by \citet{H00}.
The best scaling factor for a certain SED at each redshift
was determined by minimizing $\chi^2$.
The $\chi^2$ value was summed up at the wavelengths where the model predictions could be evaluated.
Upper limits were used as restrictions.
Thus, the number of degrees of freedom is the number of detections
(within the wavelength range of the SED)
minus one (for a scaling factor).
The smallest reduced $\chi^2$ was achieved with the Arp220 SED for sources A and B,
and with the 13\,Gyr Sc SED for source C.
The resulting $z_{\mathrm{ph}}$(SED)-values are similar to $z_{\mathrm{ph}}(6.7\,\mu\mathrm{m})$.
A spectroscopic redshift of source C
was measured as $z_{\mathrm{sp}}=1.038$
\citep{CSH+96}.
A millimetric redshift of $z_{\rm ph}\mathrm{(1.4\,GHz)}=1.8$ was derived
for source B
\citep{BCMR01}.
All of these are consistent in the range $z=1$--2
(Table~\ref{tab:z}).
Source C gives a minimum $\chi^2$ at $z \sim 1$ for many SEDs including the Arp220 SED.
The hard X-ray flux for the $Chandra$ source B
was not used in the $\chi^2$ calculations except in the case with the NGC6240 SED.
An overall fit to the hard X-ray to radio data turned out to not be so good;
the NGC6240 SED suggests $z=1.9$ but with a reduced $\chi^2=5.9$.
\citep[cf. Fig.~11 in][]{BCMR01}.
If we admit such a level of reduced $\chi^2$,
there could be some room for a higher redshift
for the optically faint sources A and B.
Source A could be at $z=5.3$ with the 0.8\,Gyr E SED (a reduced $\chi^2=1.9$).
For source B, the 0.2\,Gyr E SED gives $z=3.1$ with a reduced $\chi^2=3.2$
(for the UV to radio data only).
Both E SEDs represent a dusty elliptical before the galactic wind blows.

\section{Discussion}

We identified source C with a galaxy
offsetting 6\farcs4 from the nominal SCUBA position.
Such sizes of offsets were observed
in simulations of the source detection in a crowded field
\citep{ELW+00,H01}.
SCUBA observations of radio selected sources
actually found such offsets
at the submillimeter flux level of 2--4\,mJy with $\mathrm{S/N} = 3$--5
\citep{BCR00}.
The expected large offsets and optical faintness suggest
that the mid-infrared identification of SCUBA sources
could be an effective way to find their optical counterparts.
For the submillimeter sample presented here,
their counterparts are as faint as 14\,$\mu$Jy at 6.7\,$\mu$m, $K=20$, or $I=25$.
Integral galaxy counts at the corresponding depths
are $1\,\times\,10^4$ deg$^{-2}$,
$3\,\times\,10^4$ deg$^{-2}$, and
$8\,\times\,10^4$ deg$^{-2}$, respectively
\citep{SKC+01,MIT+01, MSC+01}.
Thus, the probability of finding spurious associations by chance
would be smaller by a factor of 3--8 at 6.7\,$\mu$m.

Our ISOCAM map taken with a pixel field of view (PFOV) of 6{\arcsec}
still has a low resolution problem.
The 6.7\,$\mu$m contour for source C
extends to the east to include an edge-on spiral.
The extended appearance is also seen for source B.
Although their fluxes were corrected to be one point source statistically
\citep{SKC+01},
effects of the extension to the corrected flux for a particular source
remain uncertain.
A mid-infrared imaging with a finer PFOV (1\arcsec--2\arcsec)
with SIRTF/IRAC or ASTRO-F/IRC
would clarify such ambiguities.
The high sensitivity of these detectors
will also strengthen redshift estimates for the milli-Jansky 850\,$\mu$m sources.
They can be detected even at $z=5$,
and their redshifts could be determined,
provided that the Arp220 SED is a representative for them.
For cases of other SEDs,
multi-band mid-infrared fluxes should be required to estimate photometric redshifts
because such high redshift sources are likely to be faint
in the optical, or even in the near-infrared.

The estimated range of redshift for the three sources
in the SSA13 field is $z$ = 1--2.
Given a cosmology with $H_0=50\,\mathrm{km}\,\mathrm{s}^{-1}\,\mathrm{Mpc}^{-1}$
and $\Omega_0=0.1$,
the sources would be 2--4 times more luminous than Arp220.
These properties are similar to other submillimeter sources
\citep{LEG+99,BCS+99,BCR00}.
A simple scaling of the model parameters for the GRASIL Arp220 SED
gives star formation rates in the range
(1--2) $\times$ $10^{3}\,M_\sun\,\mathrm{yr}^{-1}$
and stellar masses in the range (3--7) $\times$ $10^{11}\,M_\sun$.
While the peak of the dust emission was not observed directly,
the 6.7\,$\mu$m flux originated from the rest-frame near-infrared.
Because rest-frame near-infrared emission
has a good correlation with stellar mass,
it would be accurate to assume
that these submillimeter sources have massive stellar populations in them.
With signs of a major merger in all the optical counterparts,
similar to the irregular morphologies already reported
\citep{SIB+98,LEG+99},
it is suggested
that less massive galaxies identified at $z \sim 3$
\citep{PKS+98}
would evolve into local massive galaxies
via major merging with the submillimeter bright phase.

\section{Conclusions}

All three of the faint 850\,$\mu$m sources (2--4\,mJy) in the SCUBA SSA13 deep field
have been found to have ISOCAM 6.7\,$\mu$m counterparts (10--30\,$\mu$Jy).
Utilizing the smaller beam size at 6.7\,$\mu$m,
we find that all the three sources are coincident with interacting galaxy pairs
in the $HST$ $I$ band images.
They all have VLA 1.4\,GHz counterparts,
and only one of them
is detected with $Chandra$ at the 2--10\,keV band.
Based on the properties similar to local ULIGs,
we used submillimeter to mid-infrared flux ratios
to estimate photometric redshifts.
We derived other photometric redshifts
utilizing the X-ray to radio photometry data.
Both estimates are
consistent with a spectroscopic redshift and a millimetric one.
An Arp220 SED at $z=1$--2
provides a good fit to the available data.
With an assumption that AGN contributions are negligible,
the implied SFRs are in the range
(1--2) $\times$ $10^{3}\,M_\sun\,\mathrm{yr}^{-1}$
and the mid-infrared emission requires
stellar masses in the range
(3--7) $\times$ $10^{11}\,M_\sun$.

\acknowledgments

We thank Eric Richards
for permitting us to use the VLA coordinates data
prior to the publication.
We acknowledge correspondence with Dave Sanders.
We thank an anonymous referee for the comments.
The analysis has been achieved with the IDL Astronomy Users Library
maintained by Wayne Landsman.
This research has
made use of NASA's Astrophysics Data System Bibliographic Services.
A part of this research has been supported by
JSPS Research Fellowships for Young Scientists.


\begin{deluxetable}{ccccccccccc}
\tablecolumns{11}
\tablewidth{0pt}
\tabletypesize{\footnotesize}   
\tablecaption{
Identified submillimeter sources
  \label{tab:id}
  }
\tablehead{
\colhead{ID}                           &
\colhead{Name\tablenotemark{1}}        &
\colhead{RA}                           &
\colhead{Dec}                          &
\colhead{2--10\,keV\tablenotemark{2}}  &
\colhead{$B$}                          &
\colhead{$I$}                          &
\colhead{$K$}                          &
\colhead{6.7\,$\mu$m\tablenotemark{3}} &
\colhead{850\,$\mu$m\tablenotemark{1}} &
\colhead{1.4\,GHz\tablenotemark{4}}      \\
\colhead{}                                       &
\colhead{[BCS99]}                                &
\colhead{(J2000)}                                &
\colhead{(J2000)}                                &
\colhead{                                      } &
\colhead{(mag)}                                  &
\colhead{(mag)}                                  &
\colhead{(mag)}                                  &
\colhead{($\mu$Jy)}                              &
\colhead{(mJy)}                                  &
\colhead{($\mu$Jy)}
}
\startdata
A & J131232.1+424430 & 13 12 31.94 & +42 44 29.7 &       $<$3.2\phm{$7$} &       $>$25.5 &       $>$24.7 & 19.7 & 18 & 3.8 & \nodata \\
B & J131228.0+424458 & 13 12 28.29 & +42 44 54.6 & \phm{$>$}5.67         &       $>$25.5 & \phm{$>$}24.0 & 19.9 & 14 & 2.3 & 39      \\
C & J131225.7+424350 & 13 12 25.18 & +42 43 44.9 &       $<$3.2\phm{$7$} & \phm{$>$}24.3 & \phm{$>$}21.6 & 18.5 & 27 & 2.4 & \nodata
\enddata
\tablerefs{(1) \citealp{BCS99}; (2) \citealp{MCB+00}; (3) \citealp{SKC+01}; (4) \citealp{BCMR01}}
\tablecomments{
Right Ascension in hour, minute, second
and declination in degree, arcmin, arcsec
are measured at the $K$ band.
Hard X-ray fluxes are in units of 10$^{-15}\,$erg\,s$^{-1}$\,cm$^{-2}$.
For the magnitudes,
3{\arcsec} diameter aperture measurements
are corrected with average offsets
to approximate total magnitudes
\citep{CSH+96}.
Upper and lower limits are 3\,$\sigma$.
}
\end{deluxetable}

\begin{deluxetable}{ccccc}
\tablecolumns{5}
\tablewidth{270pt}
\tablecaption{
Redshifts of the submillimeter sources
  \label{tab:z}
  }
\tablehead{
\colhead{ID}                                       &
\colhead{$z_\mathrm{sp}$\tablenotemark{1}}         &
\colhead{$z_\mathrm{ph}$(SED)}                     &
\colhead{$z_\mathrm{ph}$(6.7\,$\mu$m)}             &
\colhead{$z_\mathrm{ph}$(1.4\,GHz)\tablenotemark{2}}
}
\startdata
A & \nodata & $2.0^{+0.4}_{-0.1}$ & $2.0^{+0.6}_{-0.6}$ & \nodata \\
B & \nodata & $1.6^{+0.1}_{-0.2}$ & $1.6^{+0.6}_{-0.6}$ & 1.8     \\
C & 1.038   & $0.9^{+0.1}_{-0.1}$ & $1.0^{+0.3}_{-0.3}$ & \nodata
\enddata
\tablerefs{(1) \citealp{CSH+96}; (2) \citealp{BCMR01}}
\tablecomments{
Photometric redshifts $z_\mathrm{ph}$(SED)
are derived via $\chi^2$ minimization
between the available photometry data and template SEDs.
The uncertainty ranges are
$1\,\sigma$ confidential limits
with the best fit SED.
Other estimates $z_\mathrm{ph}$(6.7\,$\mu$m)
are obtained with the Arp220 SED
(Fig.~\ref{fig:redshift}).
The $1\,\sigma$ errors originate
from $1\,\sigma$ photometry errors
at 6.7\,$\mu$m and 850\,$\mu$m.
}
\end{deluxetable}

\end{document}